\documentclass[runningheads]{llncs}
\usepackage{algorithm}
\usepackage{amsmath}
\usepackage{xurl}
\usepackage{graphicx}
\usepackage{verbatim}
\usepackage{xcolor}

\begin{document}
\title{CortexMorph: fast cortical thickness estimation via diffeomorphic registration using VoxelMorph}
%
%
\author{Richard McKinley
\and
Christian Rummel
}

\authorrunning{R McKinley and C Rummel}
%
\institute{Support Center for Advanced Neuroimaging (SCAN), University Institute of Diagnostic and Interventional Neuroradiology, Inselspital, Bern University Hospital, Bern, Switzerland}
\titlerunning{CortexMorph: CTh estimation using VoxelMorph}

\maketitle              
\begin{abstract}
    The thickness of the cortical band is linked to various neurological and psychiatric conditions, and is often estimated through surface-based methods such as Freesurfer in MRI studies. The DiReCT method, which calculates cortical thickness using a diffeomorphic deformation of the gray-white matter interface towards the pial surface, offers an
    alternative to surface-based methods. Recent studies using a synthetic cortical thickness phantom have demonstrated that the combination of \mbox{DiReCT} and deep-learning-based segmentation is more sensitive to subvoxel
    cortical thinning than Freesurfer.

While anatomical segmentation of a T1-weighted image now takes seconds, existing implementations of DiReCT rely on iterative image registration methods which can take up to an hour per volume. On the other hand, learning-based deformable image registration methods like VoxelMorph have been shown to be faster than classical methods while improving registration accuracy. This paper proposes CortexMorph, a new method that employs unsupervised deep learning to directly regress the deformation field needed for DiReCT. By combining CortexMorph with a deep-learning-based segmentation model, it is possible to estimate region-wise thickness in seconds from a T1-weighted image, while maintaining the ability to detect cortical atrophy.  We validate this claim on the OASIS-3 dataset and the synthetic cortical thickness phantom of Rusak et al.
\end{abstract}
\section{Introduction}

Cortical thickness (CTh) is a crucial biomarker of various neurological and psychiatric disorders, making it a primary focus in neuroimaging research. The cortex, a thin ribbon of grey matter at the outer surface of the cerebrum, plays a vital role in cognitive, sensory, and motor functions, and its thickness has been linked to a wide range of neurological and psychiatric conditions, including Alzheimer's disease, multiple sclerosis, schizophrenia, and depression, among others.  Structural magnetic resonance imaging (MRI) is the primary modality used to investigate CTh, and numerous computational methods have been developed to estimate this thickness  on the sub-millimeter scale.
Among these, surface-based methods like Freesurfer~\cite{fischl2012freesurfer,fischl2000measuring} have been widely used, but they are computationally intensive, making them less feasible for clinical applications.  Optimizations based on Deep Learning have brought the running time for a modified Freesurfer pipeline down to one hour.~\cite{henschel2020} The DiReCT method~\cite{das2009registration}  offers an alternative to surface-based morphometry methods, calculating CTh via a diffeomorphic deformation of the gray-white matter interface (GWI) towards the pial surface (the outer edge of the cortical band).  The ANTs package of neuroimaging tools provides an implementation of DiReCT via the function \texttt{KellyKapowski}: for readablility we refer below to \texttt{KellyKapowski} with its default parameters as ANTs-DiReCT. The ANTs cortical thickness pipeline uses ANTs-DiReCT together with a three-class segmentation (grey matter, white matter, cerebrospinal fluid) provided by the Atropos segmentation method, taking between 4 and 15 hours depending on the settings and available hardware~\cite{avants2011,tustison2013}.  A more recent version of ANTs provides a deep-learning based alternative to Atropos, giving comparable results to ANTs but accelerating the overall pipeline to approximately one hour, such that now the running time is dominated by the time needed to run ANTs-DiReCT~\cite{tustison2021}.  Meanwhile, Rebsamen et al. have shown that applying DiReCT to the output of a deep-learning-based segmentation model trained on Freesurfer segmentations (rather than Atropos) yields a CTh method which agrees strongly with Freesurfer, while having improved repeatability on repeated scans~\cite{rebsamen2020}.  Subsequently, a digital phantom using GAN-generated scans with simulated cortical atrophy showed that the method of Rebsamen et al. is more sensitive to cortical thinning than Freesurfer~\cite{rusak2022a}.

The long running time of methods for determining CTh remains a barrier to application in clinical routine:  a running time of one hour, while a substantial improvement over Freesurfer and ANTs cortical thickness, is still far beyond the real-time processing desirable
for on-demand cortical morphometry in clinical applications.  In terms of both the speed and performance, VoxelMorph and related models are known to outperform classical deformable registration methods, suggesting that a DiReCT-style CTh algorithm based on unsupervised registration models may enable faster CTh estimation.  \cite{dalca2019unsupervised,balakrishnan2019,zou2022}  In this paper, we demonstrate that a VoxelMorph style model can be trained to produce a diffeomorphism taking the GWI to the pial surface, and that this model can be used to perform DiReCT-style CTh estimation in seconds.  We trained the model on 320 segmentations derived from the IXI and ADNI datasets, and demonstrate excellent agreement with ANTs-DiReCT on the OASIS-3 dataset.  Our model also shows improved performance on the digital CTh phantom of Rusak et al.\cite{rusak2022a}

\section{Methods}
\begin{figure}[htbp]
	\centering
	\includegraphics[scale=0.20]{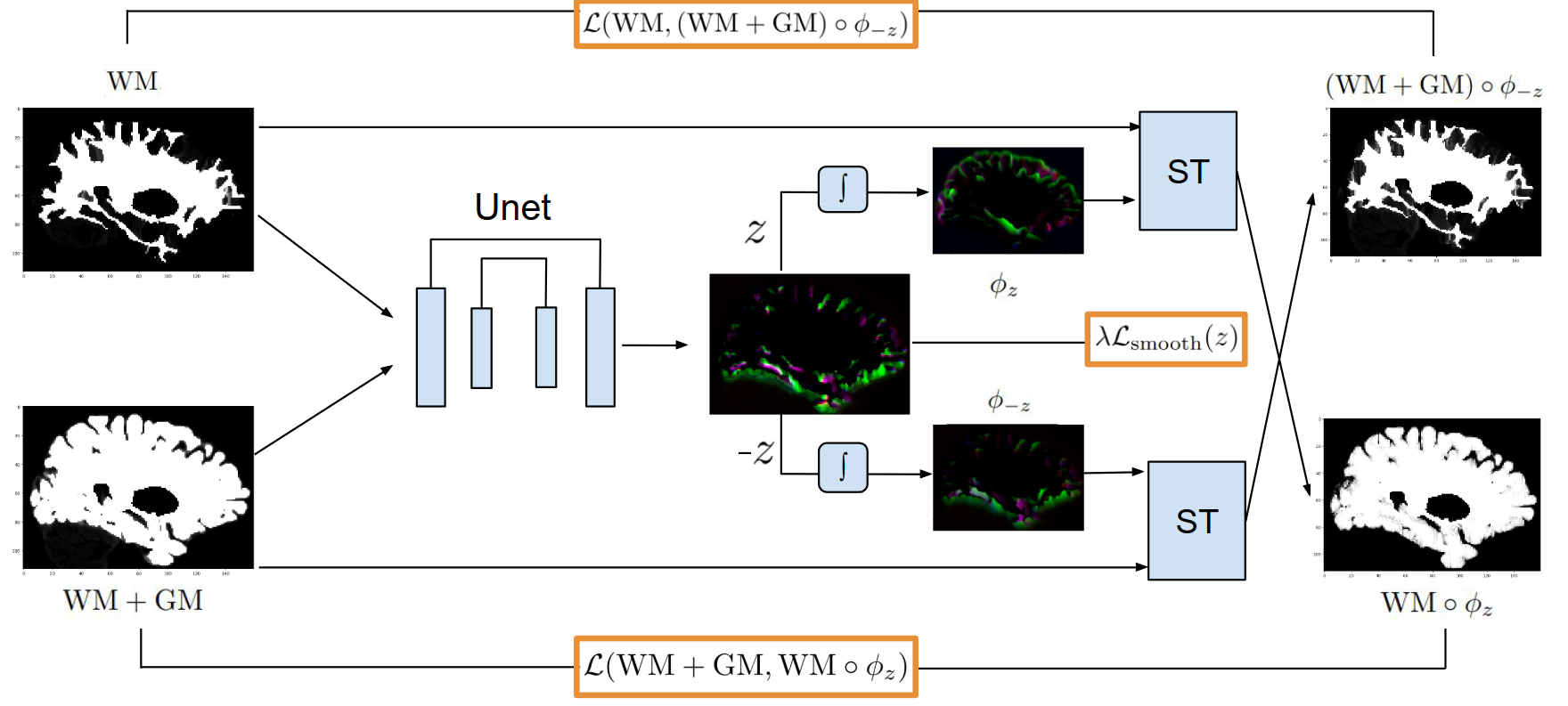}
	\caption{End-to-end unsupervised architecture for DiReCT: velocity field $z$ is regressed from WM and WM+GM segmentations, using a Unet.  This velocity field is then integrated by seven scaling and squaring layers ($\int$) to yield forward and reverse deformation fields $\phi_{z}$ and $\phi_{-z}$, which are used to deform the input images in spatial transformer (ST) blocks.  Components of the loss function are marked in orange. }
	\label{fig:arch}
\end{figure}
\subsection{DiReCT Cortical Thickness estimation}

The estimation of CTh using the DiReCT method~\cite{das2009registration} proceeds as follows: first a (partial volume) segmentation of the cortical white matter (WM) and cortical grey matter (GM) is obtained.  Second, a forward deformation field $\phi$ mapping the white-matter (WM) image towards the WM+GM image is computed. This forward deformation field should be a diffeomorphism, in order that the deformation field is invertible and the topology of the inferred pial surface is the same as the GWI.  Third, the diffeormorphism is inverted to obtain the reverse the deformation field, taking the pial surface towards the 
GWI.  Finally, the CTh is determined by computing the magnitude of the reverse field at the GWI: specifically, at each voxel of WM adjacent to the GM.   In ANTs-DiReCT, the forward transform (from WM to WM+GM) is calculated by a modified greedy algorithm, in which the WM surface is propagated iteratively in the direction of the surface normal until it reaches the outer GM surface or a predefined spatial prior maximum is reached.  The approximate inverse field is then determined by numerical means using kernel based splines (as implemented in ITK). 

The absence of a reliable gold-standard ground truth for CTh makes comparisons between methods difficult.  This situation has recently been improved
by the publication of a synthetic cortical atrophy phantom: a dataset generated using a GAN conditioned on subvoxel segmentations, consisting of 20 synthetic subjects with 19 induced sub-voxel  atrophy levels per subject (ten evenly spaced atrophy levels from 0 to 0.1mm, and a further nine evenly spaced atrophy levels from 0.1mm to 1mm).~\cite{rusak2022a}  The purpose of this digital phantom is to explore the ability of CTh algorithms to resolve subtle changes of CTh.  The paper of Rusak et al. analyzed the performance of several CTh methods on this dataset, finding that the DL+DiReCT method \cite{rebsamen2020} (which combines a deep network trained on Freesurfer annotations with ANTs-DiReCT) was the most sensitive to cortical atrophy and had the best agreement with the synthetically
induced thinning.

\subsection{CortexMorph: VoxelMorph for DiReCT}
The original VoxelMorph architecture, introduced in \cite{dalca2019unsupervised}, utilized a Unet architecture to directly regress a displacement field from a fixed brain image and a moving brain image. Application of a spatial transform layer allows the moving image to be transformed to the space of the fixed image, and compared using a differentiable similarity metric such as mean squared error or cross-correlation.  Since the spatial transformation is also a differentiable operation, the network can be trained end-to-end.  Later adaptations of the concept employed a regression of a stationary \emph{velocity field}, with the deformation field being calculated via an integration layer: the principal advantage of this formulation is that integrating through a velocity field yields a \emph{diffeomorphism}.~\cite{balakrishnan2019}  Since diffeomorphic registration is required in the DiReCT method, we adopt this velocity-field form of VoxelMorph for our purposes.

The setup of our VoxelMorph architecture, CortexMorph,  is detailed in Figure \ref{fig:arch}.  The two inputs to the network are a partial volume segmentation of white matter (WM), and a partial volume segmentation of grey matter plus white matter (WM+GM).  These are fed as entries into a Unet, the output of which is a velocity field $z$, which is then integrated using 7 steps of scaling and squaring to yield a displacement field $\phi_z$.  This displacement field is then applied to the WM image to yield the deformed white matter volume $\mathrm{WM} \circ \phi_z$.  By integrating $-z$ we obtain the reverse deformation field $\phi_{-z}$ , which is applied to the WM+GM image to obtain a deformed volume $\mathrm{(WM+GM)} \circ \phi_{-z}$.  This simplifies the DiReCT method substantially: instead of needing to perform a numerical inversion of the deformation field, the reverse deformation field can be calculated directly.  The deformed volumes are then compared using a loss function $\mathcal{L}$ to their non-deformed counterparts: both directions of deformation are weighted equally in the final objective function.  To encourage smoothness, a discrete approximation of the squared gradient magnitude of the velocity field $\mathcal{L}_\mathrm{smooth}$ is added to the loss as a regularizer.~\cite{balakrishnan2019} As a result, our loss has the following form 

\begin{equation} 
  \mathcal{L}(\mathrm{WM}, \mathrm{(WM+GM)} \circ \phi_{-z})+ \mathcal{L}(\mathrm{WM+GM}, \mathrm{WM} \circ \phi_z) + \lambda \mathcal{L}_\mathrm{smooth}(z)
  \label{eq:loss_function}
\end{equation}

\subsection{Data and WM/GM segmentation}

Training data and validation for our VoxelMorph model was derived from two publicly available sources: images from 200 randomly selected elderly individuals from the ADNI dataset \cite{jackjr.2008} and images from 200 randomly selected healthy adults from the IXI dataset (\url{brain-development.org/ixi-dataset}).  From each of these datasets, 160 images were randomly chosen to serve as training data, yielding in total 320 training cases and 80 validation cases.  For testing our pipeline, we use two sources different from the training/validation data: the well-known OASIS-3 dataset (2,643 scans of 1,038 subjects, acquired over $>10$ years on three different Siemens scanners), and the CTh phantom of Rusak et al. ~\cite{rusak2022a,RusakPhantom}

For WM/GM segmentation, we employed the DeepSCAN model~\cite{rebsamen2020,mckinley2019few}, which is available as part of DL+DiReCT (\url{https://github.com/SCAN-NRAD/DL-DiReCT}), since this is already known to give high-quality CTh results when combined with ANTs-DiReCT.    This model takes as input a T1-weighted image, performs resampling and skull-stripping if necessary (provided by HD-BET \cite{isensee2019}) and produces a partial volume segmentation $P_w$ of the white matter and $P_g$ of the cortex (the necessary inputs to the DiReCT algorithm) with 1mm isovoxel resolution.  It also produces a cortical parcellation in the same space (necessary to calculate region-wise CTh measures).  We applied this model to the training data, validation data, and the 400 synthetic MRI cases of the CTh phantom, both to produce ANTs-DiReCT CTh measurements and also as an input to our VoxelMorph models.

\subsection{Training and model selection}

Our network was implemented and trained in Pytorch (1.13.1).  We utilized a standard Unet (derived from the nnUnet framework \cite{Isensee211}) with 3 pooling steps and a feature depth of 24 features at each resolution.  The spatial transformer/squaring and scaling layers/gradient magnitude loss were incorporated from the official VoxelMorph repository.  For the loss function $\mathcal{L}$ we tested both L1 loss and mean squared error (MSE).  We tested values of the smoothness parameter lambda between $0$ and $0.05$.  The models were trained with the Adam optimizer, with a fixed learning rate of $10^{-3}$ and weight decay $10^{-5}$. Patches of size $128^3$ were used as training data in batches of size 2.  

The training regime was fully unsupervised with respect to cortical thickness: neither the deformation fields yielded by ANTs-DiReCT nor the CTh results computed from those deformation fields were used in the objective function.  Since we are interested in replacing the iterative implementation of DiReCT with a deep learning counterpart, we used the 80 validation examples for model selection, selecting the model which showed best agreement in mean global CTh with the results of ANTs-DiReCT.   The metric for agreement chosen is intraclass correlation coefficient, specifically ICC(2,1) (the proportion of variation explained by the individual in a random effects model, assuming equal means of the two CTh measurement techniques), since this method is sensitive to both absolute agreement and relative consistency of the measured quantity.  ICC was calculated using the python package \texttt{Pingouin}. \cite{Vallat2018}

\subsection{Testing}

The VoxelMorph model which agreed best with ANTs-DiReCT on the validation set was applied to segmentations of the OASIS-3 dataset, to confirm whether model selection on a small set of validation data would induce good agreement with ANTs-DiReCT on a much larger test set (metric, ICC(2,1)) and to the synthetic CTh phantom of Rusak et al, to determine whether the VoxelMorph model is able to distinguish subvoxel changes in CTh (metric, coefficient of determination ($R^2$)).

\section{Results}

\begin{figure}[htbp]
	\centering
	\includegraphics[scale=0.25]{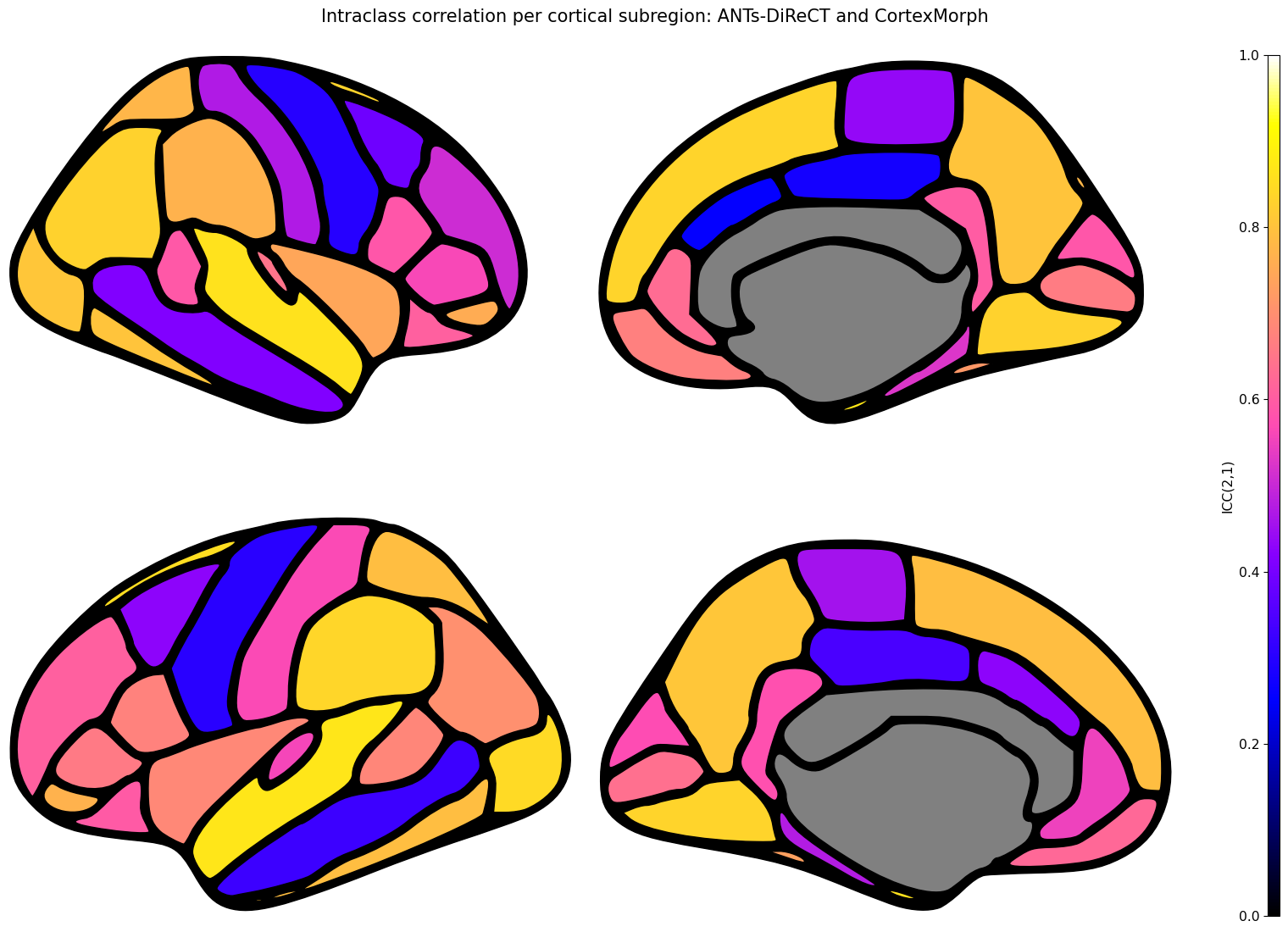}
    
	\caption{Region-wise performance of CortexMorph: ICC(2,1) of mean region-wise cortical thickness between CortexMorph and ANTs-DiReCT, using the segmentations generated by DeepSCAN on the OASIS-3 dataset.}
	\label{fig:iccs}
\end{figure}

The best performing model on the validation set (in terms of agreement with DiReCT) was the model trained with MSE loss and a $\lambda$ of 0.02.  When used to measure mean global CTh, this model scored an ICC(2,1) of 0.91 ($95\%$ confidence interval [0.9, 0.92]) versus the mean global CTh yielded by ANTs-DiReCT on the OASIS-3 dataset.  For comparison, on the same dataset the ICC between Freesurfer and the ANTs-DiReCT method was 0.50 ([$95\%$ confidence interval -0.08, 0.8]).  A breakdown of the ICC by cortical subregion can be seen in Figure~\ref{fig:iccs}: these range from good agreement  (entorhinal right, ICC = 0.87) to poor (caudalanteriorcingulate right, ICC=0.26), depending on the region.  However, ICC(2,1) is a measure of absolute agreement,  
as well as
correlation: all regional Pearson correlation coefficients lie in a range [0.64-0.90] (see supplementary material for a region-wise plot of the Pearson correlation coefficients).

\begin{figure}[htbp]
	\centering
	\includegraphics[scale=0.50]{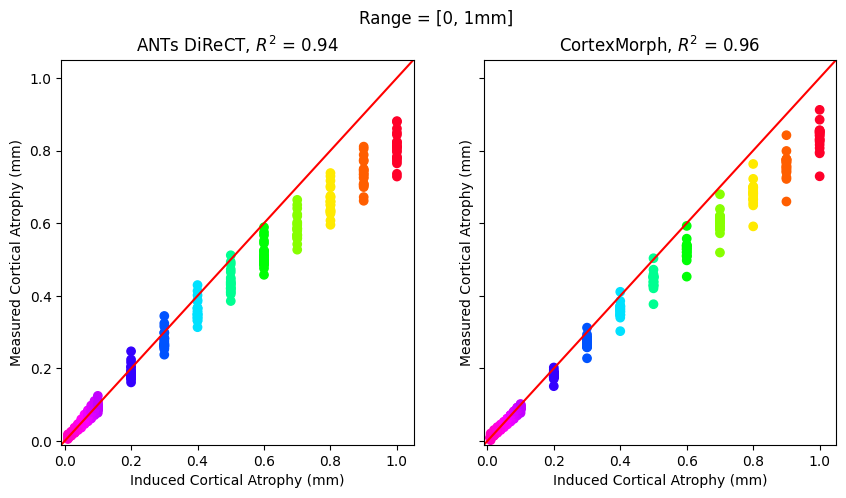}
    \includegraphics[scale=0.50]{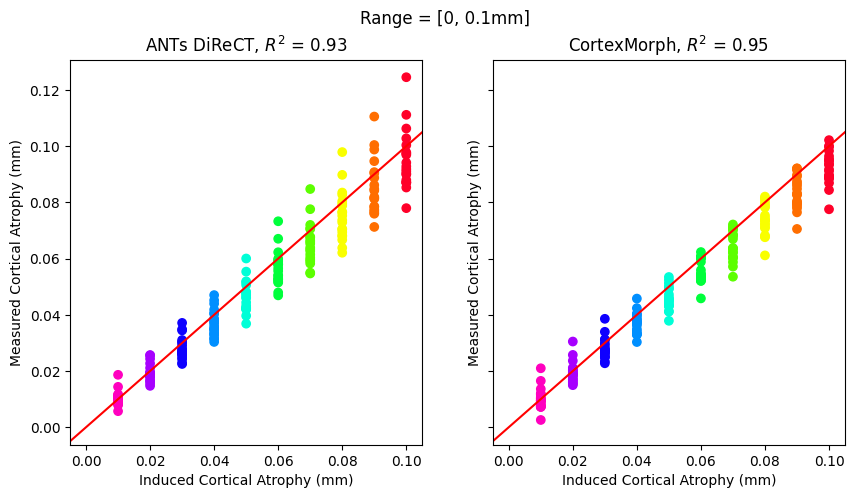}
	\caption{Performance of ANTs-DiReCT and CortexMorph on the CTh phantom of Rusak et al, based on segmentations derived from DeepSCAN.  Above: performance on the whole synthetic dataset, comprising twenty synthetic individuals, each with a baseline scan and 19 'follow-up' images with induced levels of uniform cortical atrophy.  Measured atrophy is defined as the difference between the mean CTh as measured on the synthetic baseline scan and the mean CTh measured on the synthetic follow-up, averaged across the whole cortex.  Below: The same data, but focused only on the range [0-0.1mm] of induced atrophy.  $R^2$ denotes the coefficient of determination between the induced and measured atrophy levels.}
	\label{fig:results}
\end{figure}

Performance of this model on the CTh digital phantom can be seen in Figure~\ref{fig:results}: agreement with the induced level of atrophy is high (metric: Coefficient of Determination between the induced and the measured level of atrophy, across all 20 synthetic subjects) in both the wide range of atrophy (up to 1mm) and the fine-grained narrower range of atrophy (up to 0.1mm), suggesting that the VoxelMorph model is able to resolve small changes in CTh.

Calculating regional CTh took between 2.5s and 6.4s per subject (mean, 4.3s, standard deviation 0.71s) (Nvidia A6000 GP, Intel Xeon(R) W-11955M CPU).

\section{Conclusion}
Our experiments suggest that the classical, iterative approach to cortical thickness estimation by diffeomorphic registration can be replaced with a VoxelMorph network, with $\sim$ 800 fold reduction in the time needed to calculate CTh from a partial volume segmentation of the cortical grey and white matter.  Since such segmentations can also be obtained in a small number of seconds using a CNN or other deep neural network, we have demonstrated for the first time reliable CTh estimation running on a timeframe of seconds.  This level of acceleration offers increased feasibility to evaluate CTh in the clinical setting.  It would also enable the application of ensemble methods to provide multiple thickness measures for an individual: given an ensemble of, say, 15 segmentation methods, a plausible distribution of CTh values could be reported for each cortical subregion within one minute: this would allow better determination of the presence of cortical atrophy in an individual than is provided by point estimates.  We are currently investigating the prospect of leveraging the velocity field to enable fast calculation of other morphometric labels such as grey-white matter contrast and cortical curvature: these too could be calculated with error bars via ensembling.

This work allows the fast calculation of diffeomorphisms for DiReCT on the GPU.  We did not consider the possibility of directly implementing/accelerating the classical DiReCT algorithm on a GPU in this work.  Elements of the ANTs-DiReCT pipeline implement multithreading, yielding for example a 20 minute runtime with 4 threads: however, since some parts of the pipeline cannot be parallelized it is unlikely that iterative methods can approach the speed of direct regression by CNN.

Given the lack of a gold standard ground truth for CTh, it is necessary when studying a new definition of CTh 
to compare to an existing silver standard method: this would typically be Freesurfer, but recent results suggest that this may not be the optimal method when studying small differences in CTh.  \cite{rusak2022a}
We have focused on comparison to the DL+DiReCT method for this study, since the results of this model on the CTh phantom are already reported and represent the state-of-the-art.  For this reason, it made sense to use the outputs of the underlying CNN as inputs to our pipeline.  However, the method we describe is general and could be applied to any highly performing segmentation method. Similarly, while we performed model selection to optimize agreement with the CTh values produced by Rebsamen et al, this optimization could easily be tuned to instead optimize agreement with Freesurfer.  Alternatively, we could abandon agreement and instead select models based on consistency (given by a different variant of ICC) or Pearson correlation with a baseline model: this could lead to models which deviate from the baseline model but are better able to capture differences between patients or cohorts.

\section*{Acknowledgements}

This work was supported by a Freenovation grant from the Novartis Forschungsstiftung, and by the Swiss National Science Foundation (SNSF) under grant number 204593 (ScanOMetrics).

\label{sec:disc_conclusion}

\bibliographystyle{splncs04}
\bibliography{MICCAI2023.bib}

\end{document}


\title{CortexMorph: fast cortical thickness estimation via diffeomorphic registration using VoxelMorph:supplementary material}
%
%
\author{Richard McKinley
\and
Christian Rummel
}

\authorrunning{R McKinley and C Rummel}
%
\institute{Support Center for Advanced Neuroimaging (SCAN), University Institute of Diagnostic and Interventional Neuroradiology, Inselspital, Bern University Hospital, Bern, Switzerland}
\titlerunning{CortexMorph: Supplementary material}

\maketitle              
\begin{figure}[htbp]
	\centering
	\includegraphics[scale=0.25]{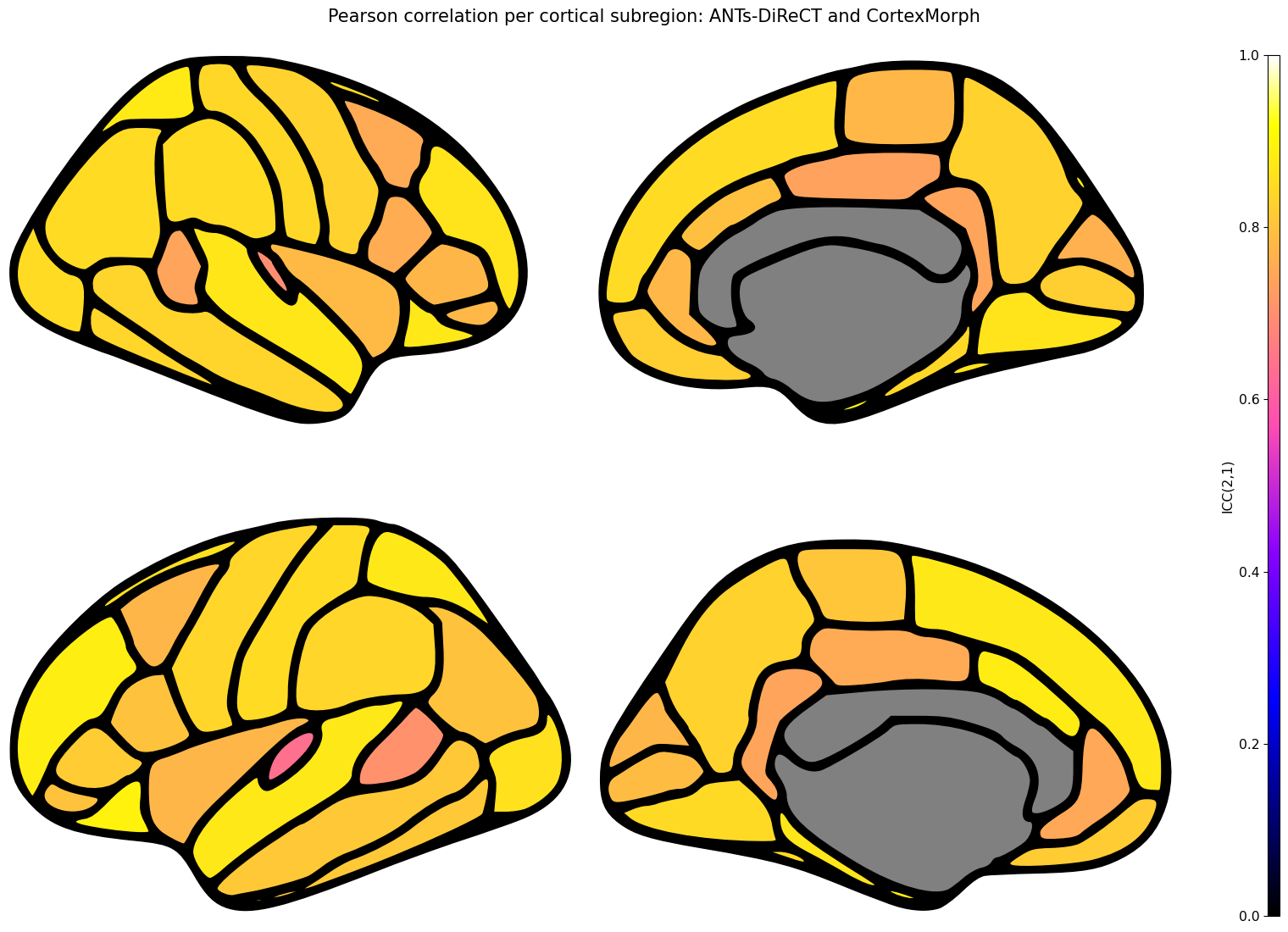}
    
	\caption{Region-wise performance of CortexMorph: Pearson correlation of mean region-wise cortical thickness between CortexMorph and ANTs-DiReCT, using the segmentations generated by DeepSCAN on the OASIS-3 dataset.} 
	\label{fig:pearson}
\end{figure}

\begin{figure}[htbp]
	\centering
	\includegraphics[scale=0.25]{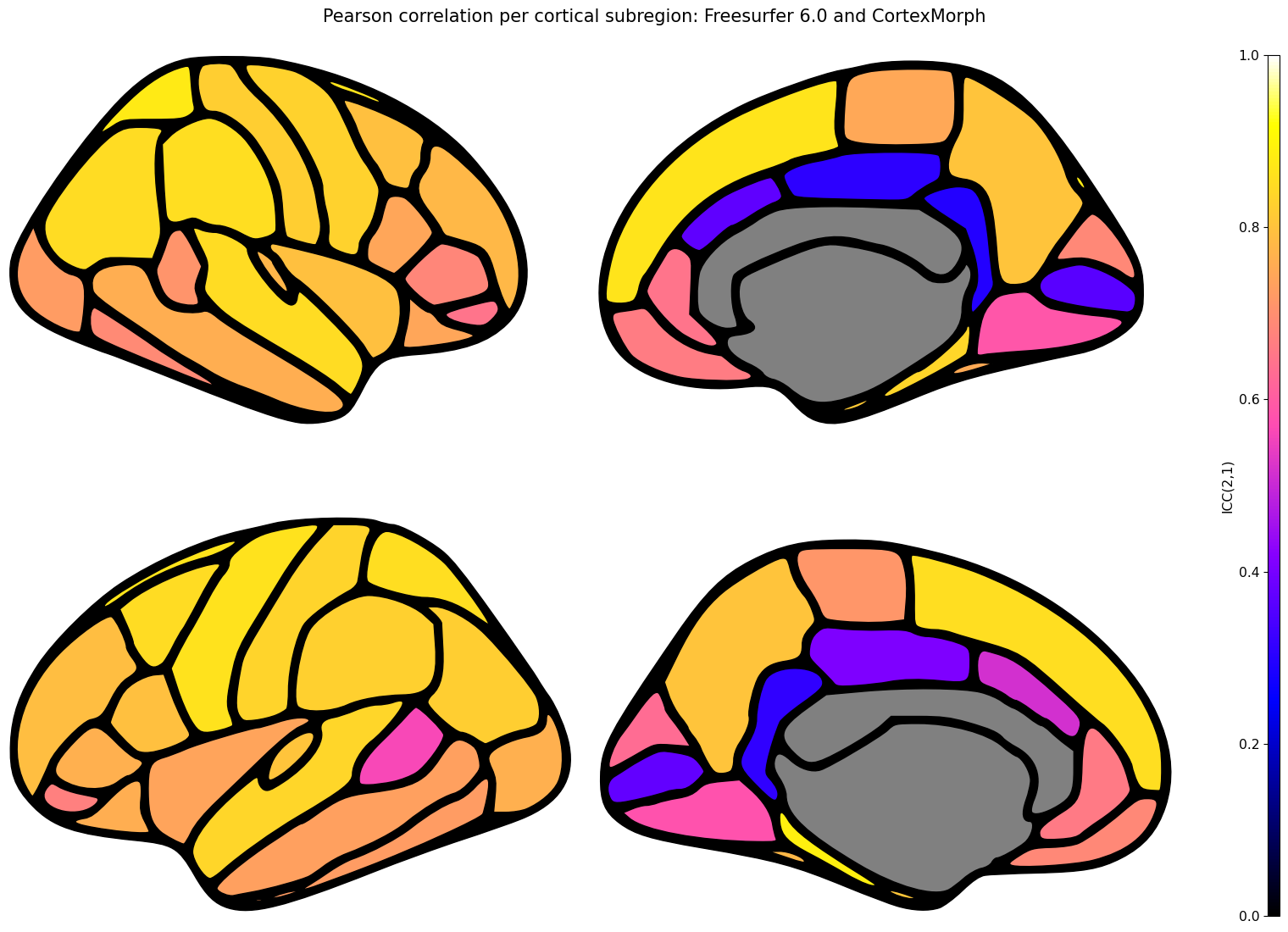}
    
	\caption{Region-wise agreement of CortexMorph versus Freesurfer : Pearson correlation of mean region-wise cortical thickness between CortexMorph and Freesurfer 6.0 on the OASIS-3 dataset.} 
	\label{fig:pearson}
\end{figure}

\begin{figure}[htbp]
	\centering
	\includegraphics[scale=0.25]{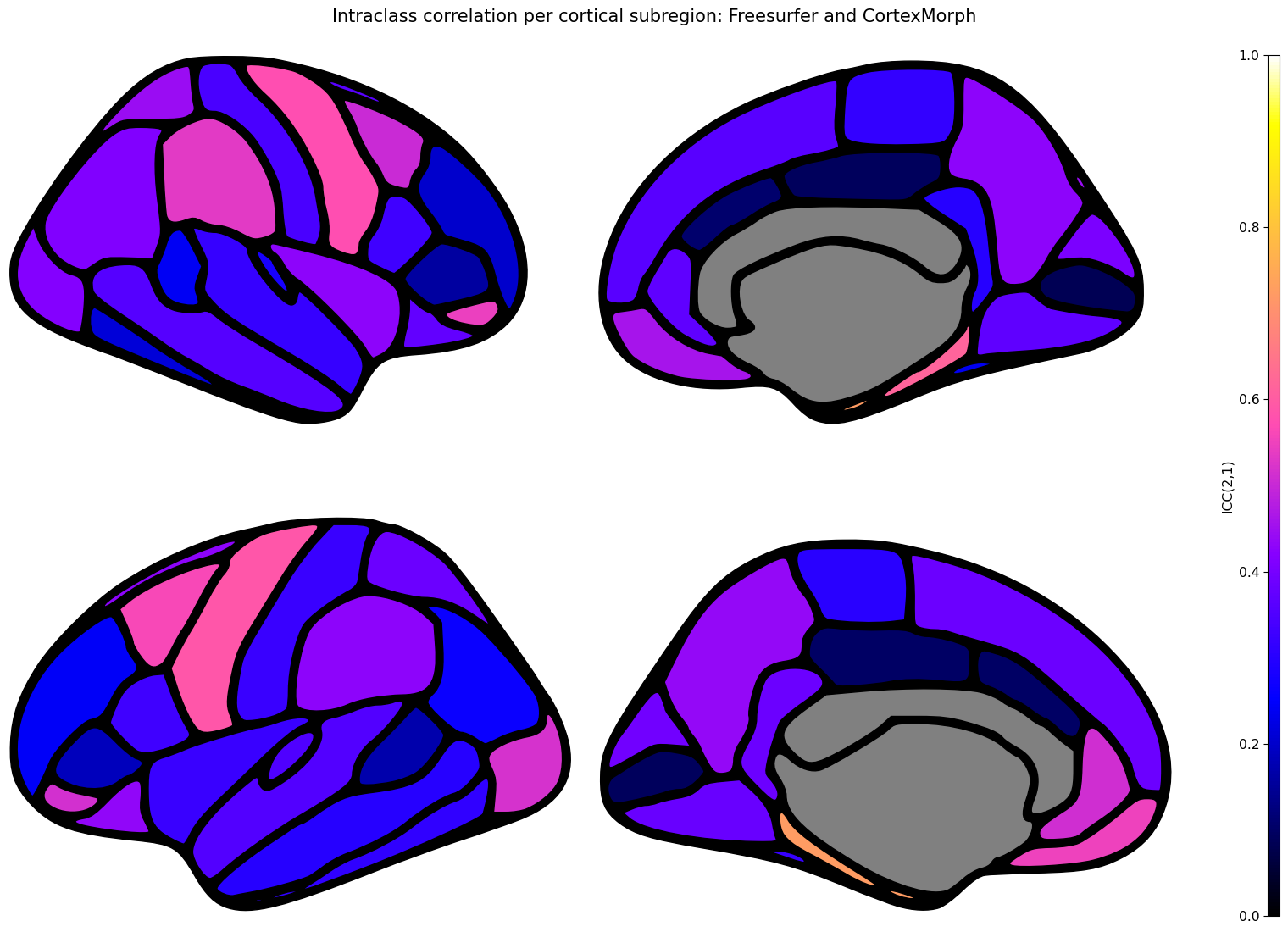}
    
	\caption{Region-wise performance of CortexMorph versus Freesurfer : ICC(2,1)  of mean region-wise cortical thickness between CortexMorph and Freesurfer 6.0  on the OASIS-3 dataset.} 
	\label{fig:pearson}
\end{figure}

\begin{table}[]
\centering
\begin{tabular}{ccccll}
                           &                          & \multicolumn{2}{c}{LossFunction } &  &  \\
                           & \multicolumn{1}{c|}{}    & L1          & MSE        &  &  \\ \cline{2-4}
\multirow{5}{*}{$\lambda$} & \multicolumn{1}{c|}{0}   & 0.63        & 0.01           &  &  \\
                           & \multicolumn{1}{c|}{0.01} & 0.54        & 0.89       &  &  \\
                           & \multicolumn{1}{c|}{0.02} & 0.97        & 0.98       &  &  \\
                           & \multicolumn{1}{c|}{0.03} & 0.96        & 0.96       &  &  \\
                           & \multicolumn{1}{c|}{0.05} & 0.72        & 0.94       &  & \\
\end{tabular}

\caption{Sensitivity of the model selection criterion (ICC(2,1) of the mean cortical thickness between the model and ANTs-DiReCT) for several values of the smoothing parameter $\lambda$, and for both L1 and MSE loss} 
\end{table}